\documentclass[letter]{aa} 
\usepackage{graphicx}
\usepackage{natbib}
\usepackage{longtable,lscape}
\usepackage{txfonts}
\begin{document}

\title{Discovery of thermal X-ray emission in the Supernova Remnant G337.8$-$0.1 (Kes 41)}

\author{J.~A. Combi\inst{1,3}, J.F. Albacete-Colombo\inst{2,3}, J. Mart\'{\i}\inst{1}
}
\authorrunning{Combi et~al.}
\titlerunning{XMM-{\it Newton} observations of G337.8$-$0.1} 
\offprints{J.A. Combi}

\institute{Departamento de F\'{\i}sica (EPS), Universidad de Ja\'en,
Campus Las Lagunillas s/n, A3, 23071 Ja\'en, Spain\\
\email{[jcombi:jmarti]@ujaen.es}
\and
Centro Universitario Regional Zona Atl\'antica (CURZA). Universidad Nacional del COMAHUE, Monse\~nor Esandi y Ayacucho (8500),
Viedma (Rio Negro), Argentina.\\
\email{donfaca@gmail.com}
\and
Facultad de Ciencias Astron\'omicas y Geof\'{\i}sicas, Universidad Nacional de La Plata, Paseo del Bosque, B1900FWA La Plata, Argentina.\\
             }

\date{Received; accepted }

 
  \abstract {} {We report here on the first detection at X-ray
wavelengths of the Supernova Remnant (SNR) G337.8$-$0.1, carried out
with the XMM-{\it Newton} Observatory.} {Using the X-ray observations,
we studied the X-ray morphology of the remnant at different energy
ranges, analysed the spectral properties and investigated a possible
variable behavior.} {The SNR shows a diffuse filled-center structure
in the X-ray region with an absence of a compact source in its center. We
find a high column density of $N_{\rm H} >$ 6.9$\times$10$^{22}$
cm$^{-2}$, which supports a relatively distant location ($d \geq$ 7
kpc). The X-ray spectrum exhibits emission lines, indicating that the
X-ray emission has a thin thermal plasma origin, and is well
represented by a non-equilibrium ionization (NEI) plasma model. The
X-ray characteristics and well-known radio parameters show that
G337.8$-$0.1 belongs to the emerging class of mixed-morphology (MM)
SNRs.} {}


\maketitle
%

\section{Introduction}

G337.8$-$0.1 (also known as Kes 41) was discovered with the Molonglo
telescope at 408 MHz (Shaver \& Goss 1970). This object, located in the
southern galactic plane, was reported in the Molonglo Observatory
Synthesis Telescope (MOST) Catalog (Whiteoak \& Green 1996) as a
bright small-diameter SNR and distorted shell. The remnant is
6'$\times$9' in size, distinctly elongated in the northeast-southwest
direction. Based on HI observations, Caswell et al. (1975) placed the
SNR beyond the tangent point at 7.9 kpc. Koralesky et al. (1998)
detected maser emission in the object at $-45$ km s$^{-1}$, implying a
far kinematic distance of 12.4 kpc. This fact is characteristic of a
shock interaction with dense molecular gas (Green et al. 1997).  The
MOST flux density of S$_{\rm 843 MHz}$=18 Jy, combined with S$_{\rm
408 MHz}$=26 Jy (Shaver \& Goss 1970), gives a spectral index of
$\alpha \sim -0.51$ for the remnant. 

At X-ray frequencies the object was never detected by early X-ray
missions. This non-detection is likely a result of the absorption of
the typically soft (i.e., $<$ 2 keV) X-ray emission by the high 
column density of gas and dust in the Galactic plane. As a part of an
effort to identify  and study the X-ray emission of the SNR
G337.8$-$0.1 we have used the greatly enhanced X-ray sensitivity
provided by the XMM-{\it Newton} telescope. An instrument with these
characteristics offers a unique opportunity to study heavily obscured
SNRs.

In this {\it Letter} we report the first detection at X-ray
wavelengths of SNR G337.8$-$0.1. The object is characterized by an
apparent centrally brightened X-ray morphology and a spectrum which
suggests that the bright central emission is thermal in nature. The
structure of the paper is as follows: in Sec.~\ref{observations} we
describe the XMM-{\it Newton} observations and data reduction. X-ray
analysis and results are presented in Sec.~\ref{analysis}. In 
Sec.~\ref{discusion} we discuss the implications of our results and
summarize the main conclusions. 

\section{Observations and data reduction} \label{observations}

The SNR G337.8$-$0.1 was observed on September 2005 by the
XMM-{\it Newton} X-ray satellite (Obs-Id. 0303100101).  The
observation was centered towards the source ($\alpha_{\rm J2000.0}$=
16$^{\rm h} 39^{\rm m} 06\fs5$, $\delta_{\rm J2000.0}=-46\degr
57\arcmin 58\farcs0$), and images were acquired with the EPIC MOS
\citep{tur01} and EPIC PN \citep{stru01} cameras. The observation was
made through a "{\it thin}" filter, and in the full frame (FF) imaging
mode. The observation was retrieved from the XMM-Newton Science
Archive (XSA)\footnote{http://xmm.vilspa.esa.es/xsa/}, and raw EPIC
data were calibrated using the last version of the Standard Analysis
System 
(SAS-V7.1.2)\footnote{http://xmm.vilspa.esa.es/external/xmm\_sw\_cal/sas.shtml}.
To construct images, spectra and light curves, we selected events with
{\sc flag==0}, and {\sc patterns}$\leq$ 12 and 4 for the MOS and PN
cameras, respectively. 

\begin{figure}[t!] 
\resizebox{1.0\hsize}{!}{\includegraphics[angle=0]{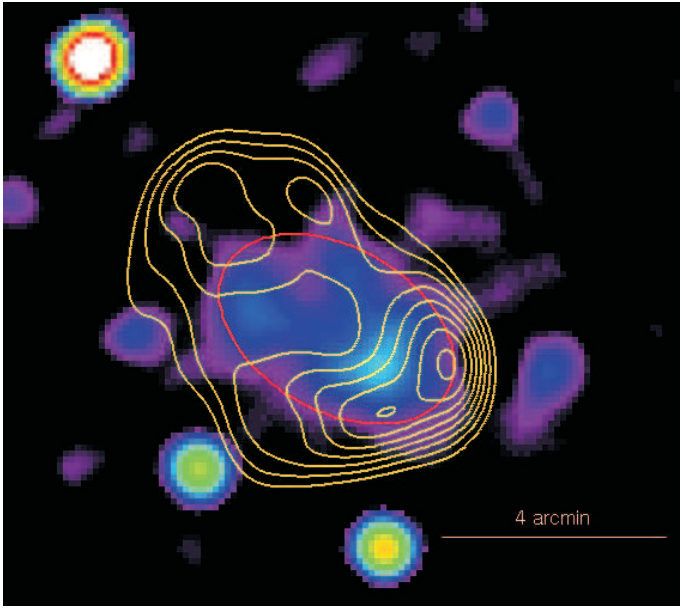}} \caption{XMM-{\it Newton} image of Kes 41 in the 0.5-10 keV energy range. The radio map observed with the MOST at 843 MHz is overlapped in yellow contours (Whiteoak \& Green, 1996). Radio contours are indicated in steps of 0.2, 0.29, 0.38, 0.47, 0.56, 0.65, 0.74 and 0.83 mJy beam$^{-1}$. The resolution is $\sim$ 43"$\times$43" cosec ($\delta$).  The red ellipse shows the region from which the X-ray spectra was extracted (see Sec.3.2).} 
\label{fig:imgX1}
\end{figure}

We derive Good Time Intervals (GTI)  by the accumulation of background
light-curves in the 10-15 keV energy band, which leads to a reduction
of $\sim$15\% in the net exposure time of the observation. The net
exposure time of the observation for all EPIC cameras finally reaches
about 44.8 ks. Unfortunately, the object was negatively
affected by bad columns and CCD gaps in the EPIC-PN camera, which
makes the spatial and spectral study of the X-ray emission
from the SNR unreliable. In order to avoid systematics that affect the
reliability of our analysis, we made use of only the EPIC-MOS1/2
data.  

\section{X-ray analysis of G337.8$-$0.1} \label{analysis}
\subsection{Image}

We use the clean event files to generate MOS1 and MOS2 images in the
energy band [0.5-10] keV with a spatial binning of 6.53 arcsec per
pixel. In order to increase the signal-to-noise (S/N) ratio, we use
the {\it emosaic} SAS task to merge the two MOS1/2 images.
The corresponding set of exposure maps for each camera was 
prepared to account for spatial quantum efficiency and mirror
vignetting  by running the SAS task {\it eexmap}. Exposure vignetting
corrections were performed by dividing the superposed count image by
the corresponding superposed exposure maps. We adaptively smoothed
this image to a S/N ratio of 10 using the SAS task {\it asmooth}. 

Figure 1 shows the X-ray image of the SNR G337.8$-$0.1 in the 0.5-10.0
keV energy band with the radio contours at 843 MHz superposed. The
image reveals diffuse X-ray emission with an apparent filled-center
structure and the absence of a compact source in its center. The X-ray
peak of the SNR G337.8$-$0.1 is located at ($\alpha_{\rm
J2000.0}=16^{\rm h} 38^{\rm m} 55\fs7$, $\delta_{\rm J2000.0}=-46\degr
58\arcmin 32\farcs4$), and as can be seen, the X-ray emission region is
smaller than the radio structure. In fact, the X-ray emission fills
the interior of the radio shell and lies mainly in the southwest part
of the radio shell.

To examine the morphology of the remnant in more detail, we have
generated narrow-band images in the energy ranges 0.5-2 keV, 2-4 keV,
4-6 keV and 6-10 keV. In Figure 2 we show the images in
the four narrow energy bands. The source is almost undetected below 2
keV. The distribution of the emission in the 2-6 keV range is quite
similar to the broadband image. The centroid of the hard X-ray
emission lies in the SW region of the remnant. The absorbed X-ray
fluxes for the bands are: F$_{\rm 0.5-2 keV}$=1.3$\times$10$^{-14}$
erg\,s$^{-1}$\,cm$^{-2}$, F$_{\rm 2-4 keV}$ = 1.6$\times$10$^{-13}$
erg\,s$^{-1}$\,cm$^{-2}$, F$_{\rm 4-6 keV}$=7.35$\times$10$^{-14}$
erg\,s$^{-1}$\,cm$^{-2}$  and F$_{\rm 6-10 keV}$=2.9$\times$10$^{-14}$
erg\,s$^{-1}$\,cm$^{-2}$, which correspond to 5\%, 58\%, 26\% and 11\%
of the total X-ray emission, respectively. As can be seen, mostly of
the X-ray emission originates in the 2-6 keV band.   

\subsection{Spectral analysis} 

\begin{figure}[t!] 
\resizebox{1.0\hsize}{!}{\includegraphics[angle=0]{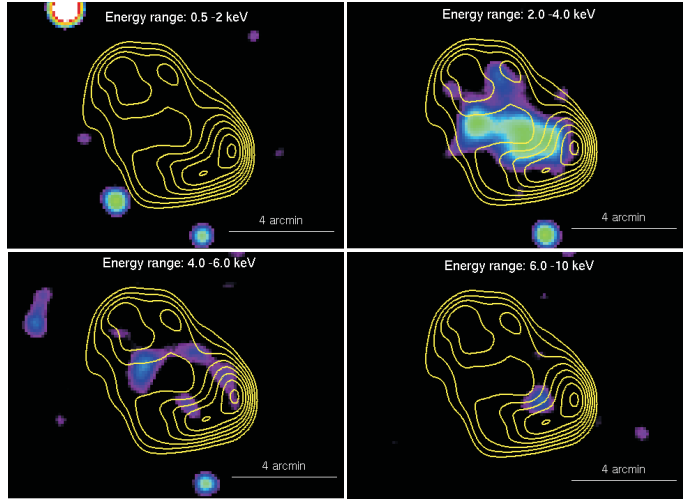}}
\caption{Composite XMM-{\it Newton} images of G337.8$-$0.1 at low,
median and high energy (0.5-2 keV, 2-4 keV, 4-6 keV, 6-10 keV) for the
combined MOS1/2 cameras. Note the lack of soft X-ray photons in the
0.5-2 keV energy range, while mostly of the X-ray emission is clearly
observed in the 2-4 keV image and the semi-circular structure in the 4-6 keV band.}  \label{fig:imgX1}
\end{figure}

For the spectral analysis we used MOS1/2 data. It was performed using
the {\sc xspec} package \citep{arnaud96}. Since the statistics of the
source are not complete enough to perform a spatially resolved spectral
analysis, we extracted X-ray photon events for the whole source by
using an elliptical region with a major axis of 5 arcmin and a minor
axis of 3.2 arcmin, as indicated in Fig.1. The background region was
taken from a nearby blank region in the neighborhood of the source.
Ancillary response files (ARFs) and redistribution matrix files (RMFs)
were calculated. The spectra were grouped with a minimum of 12 counts
per bin. The background-subtracted spectra of the MOS data are shown
in Figure 3. At high X-ray energies (above 6 keV), the spectra show 
features with low statistical significance ($\sim$ 1 to 1.5 sigma),
and are probably related to fluorescence lines in the background
spectrum of the XMM-{\it Newton} \citep[e.g.][]{DeLuca04}. 

\begin{figure}[t!] 
\resizebox{\hsize}{!}{\includegraphics[angle=0]{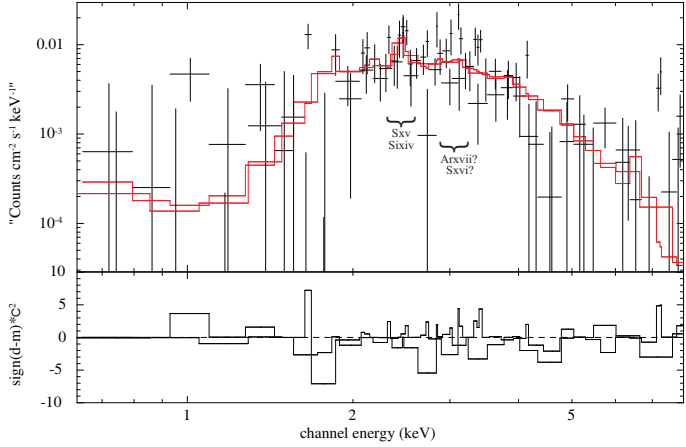}} 
\caption{The XMM-{\it Newton} MOS1/2 spectrum of the diffuse X-ray
emission of Kes 41. The solid line indicates the best fit for the NEI
model (see Table 1). The soft X-ray excess are probably due to
foreground emission. The position of the most intense transition lines
of S{\sc xv} and Si{\sc xiv} elements are indicated. The total number
of photons is 3725 and 2539 for MOS1 and MOS2, respectively.} 
\label{fig:integra}  \end{figure}

Our analysis of the XMM-{\it Newton} EPIC spectra was 
performed using a NEI model (Mazzotta et al. 1998) (see Fig. 3).
The model was affected by a multiplicative absorption {\sc wabs}
({\sc wabs$_{\rm ISM}$}\,+\,{\sc wabs$_{\rm SNR}$}) model
\citep[][]{morri83}. The goodness of the model fit was derived
according to the $\chi^2$-test statistics. C-statistics was also
applied to the spectral fit, giving similar results. Results of the
spectral analysis are summarized in Table 1. The presence of excess
soft emission is not surprising, given that Kes 41 is located in a
crowded region of the galaxy at a distance of between 7 to 12 kpc.
Emission-line features corresponding to the blend of S\,{\sc xv} and
Si\,{\sc xiv} are clearly visible at $\sim$ 2.4 keV. Possible
detection of others elements like  Ar\,{\sc xvii} and S\,{\sc xvi}
emission lines appear blended around 3.1 keV. Although the presence of
these lines support a thermal nature for the X-ray emission, the
absence of significant Fe\,{\sc xxv} at 6.7 keV does not allow us to
discard a weak non-thermal contribution at high X-ray energies. 

\begin{table}
\label{model}
\begin{center}
\caption{X-ray spectral parameters of the SNR G337.8-0.1}
\begin{tabular}{lcc}
\hline
Model:		& \multicolumn{2}{c}{NEI}\\
\cline{2-3}
Parameter 	&	Value	& 1$\sigma$ error \\
\hline
N$_{\rm H}^{\rm ISM}$ [cm$^{-2}$]      & 2.2$\times$10$^{22}$ &   fixed  \\
N$_{\rm H}^{\rm SNR}$ [cm$^{-2}$]      & 4.7$\times$10$^{22}$ &   1.0$\times$10$^{22}$     \\
kT [keV]                    & 1.4 &   0.2     \\
Abundance$^\dag$            & 0.6 &   0.3     \\
log($\tau$) [s\,cm$^{-3}$]  & 11.0&   0.3    \\
Normalization	            & 3.4$\times$10$^{-3}$ &   1.1$\times$10$^{-3}$    \\
Flux [erg\,cm$^{-2}$\,s$^{-1}$] & 7.5$\times$10$^{-12}$ &  1.8$\times$10$^{-12}$   \\
$\chi^2$/$\nu$-d.o.f.       & 1.08/380&   $--$ \\
\hline
\end{tabular}
\end{center}

Notes: The ISM H{\sc i} absorption column density (N$_{\rm H}^{\rm
ISM}$) was computed by following Dickey \& Lockman (1990) HI in the
Galaxy, and kept fixed in the fit procedure. The logarithm of
ionization timescale log($\tau$) is also written as log(n$_e$t). Flux
is absorption-corrected in the 0.5-10.0 keV energy range.
Normalization was calculated according to:
10$^{-14}$/4$\pi\,r^2\int\,n_e\,n_H dV$. $\dag$ The abundance
was adopted from \citet{anders89} and left as a free parameter in all
our fits.  \end{table} 

To obtain a statistical assessment of the X-ray variability of the SNR
G337.8$-$0.1, we use the 44.8 ks EPIC-PN observation to compare
the time arrival distribution of source photons by means of the
Kolmogorov-Smirnov (KS) test \citep{press92}. We use an extraction
region centered on the whole X-ray  emission. We see no significant
pulsed signal with a period greater than twice the read-out time of
the EPIC-PN camera in the FF mode (73.3 ms), which
corresponds to a Nyquist limit of 0.146 s. At first glance, the
absence of any significant X-ray variability and a thermal origin for
the X-ray emission allows us to discard an energetic pulsar associated with 
the SNR.  

\section{Discussion} \label{discusion}

The SNR G337.8$-$0.1 is characterized by an apparent centrally filled
X-ray morphology not well correlated with the radio shell, and
thermally dominated X-ray emission. Thus, Kes 41 could be classified
as a thermal composite (or mixed-morphology) remnant. Moreover, at
radio frequencies, the emission is non-thermal (Whiteoak \& Green
1996) and can be interpreted as synchrotron radiation from accelerated
high-energy electrons at the SNR shock. These characteristics place
the remnant in the emerging class of SNR that includes objects such as
W44 (Jones et al. 1993), 3C 391 (Rho \& Petre 1996); G272.2$-$3.2
(Harrus et al. 2001) and G290.1$-$0.8 (Slane et al. 2002). A list
containing SNRs with these characteristics was proposed by Rho \&
Petre (1998).

Although the evolutionary states that led to the observed properties
in these SNRs is not well understood, it is widely believed that these
peculiar characteristics are linked to inhomogeneities in the ISM.
Several possible scenarios have been introduced in the past to
explain thermal X-ray radiation inside radio shells of SNRs, i)
cloudlet evaporation in the SNR interior (White \& Long 1991), ii)
thermal conduction smoothing out the temperature gradient across the
SNR and enhancing the central density (Cox et al. 1999), iii) a
radiatively cooled rim with a hot interior (Harrus et al. 1997), and
iv) possible collisions with molecular clouds (Safi-Harb et al. 2005).

G337.8$-$0.1 lies in a complex region of the sky, possibly
adjacent to a very massive giant molecular cloud (Dame et al. 1986)
and close to the associated maser emission detected at $-45$ km
s$^{-1}$ by Koralesky et al. (1998). If the SNR is indeed propagating
in a region with a large number of clouds, the most suitable scenario
for describing the SNR evolution is that developed by White \& Long
(1991). In such remnants, the clouds evaporate by saturated conduction
(Cowie \& McKee, 1977) and the X-ray morphologies may be very
different from ordinary shell-like SNRs. 

Based on the X-ray properties of G337.8$-$0.1, specifically the
filled-center morphology and detected lines in the spectrum, a thermal
interpretation of the X-ray emission is the most plausible to describe 
the SNR evolution. Using the X-ray image of G337.8$-$0.1 we estimated
the volume $V$ of the X-ray emitting plasma, assuming that the plasma
fills an ellipsoid similar to the region from which the X-ray spectra
was extracted (with a diameter of 5'$\times$3'.2). For a distance
range from 7 to 12 kpc, we then obtained a volume of
(2.8-14.8)$\times$10$^{58}$ cm$^{3}$. Based on the emission measure (EM)
determined by the spectral fitting, we can estimate the electron
density of the plasma, n$_{e}$, by n$_{e}$=$\sqrt{EM/V}$, which varies 
between 0.20 cm$^{-3}$ and 0.26 cm$^{-3}$. In this case, the number
density of the nucleons was simply assumed to be the same as that of
electrons. The range of age $t$ was then determined from the
ionization timescale, $\tau$, by $t$=$\tau$/n$_{e}$. Therefore, the
elapsed time after the plasma was heated is within (12.000-16.000) yr. 
The total mass of the plasma $M_{total}$ was
estimated by $M_{total}$=n$_{e}$V\,m$_{\rm H}$ = (5\,-\,32)
M$_{\odot}$, where m$_{\rm H}$ is the mass of a hydrogen atom. These
results are consistent with the expected values for a
middle-age SNR.

One of the most detailed studies of middle-age and center-filled
thermal emission in a SNR interacting with a dense molecular cloud is
the W44 model (Cox et al. 1999; Shelton et al, 1999; Shelton et al.
2004). The proximity to molecular clouds of this SNR suggests that the
environmental interstellar matter, where the SNR is propagating, is
relatively dense. Using Cox's model we have roughly estimated some
dynamical characteristics (i.e., the shock radius $R_{\rm s}$ and
shell velocity $v_{\rm shell}$) for G337.8$-$0.1. Assuming an energy
for the explosion of $\sim$ 10$^{51}$ ergs, we found that $R_{\rm s}$=
(8-10) pc and $v_{\rm shell}$= (180-200) km s$^{-1}$. These values are
consistent with the angular size and range of distance for the SNR under
consideration. Thus, G337.8$-$0.1 could be, a SNR with characteristics similar to W44.

Finally, we have compared the X-ray luminosity of G337.8$-$0.1 with
other proto-typical mixed-morphology SNRs. The unabsorbed X-ray
luminosities in the 0.5-10 keV band of G337.8$-$0.1 for a range of
distance between 7 and 12 kpc is L$_{\rm X} \sim$ 4.5$\times$10$^{34}$
erg/s and 1.3$\times$10$^{35}$ erg/s, respectively. The first value is
similar to the X-ray luminosity of W28 (L$_{\rm
X}$$\sim$6$\times$10$^{34}$ erg s$^{-1}$; Rho \& Borkowski, 2002),
G272.2-3.2 (L$_{\rm X} \sim$ 5.2$\times$10$^{34}$ erg s$^{-1}$; Harrus
et al. 2001), and Kes 32 (L$_{\rm  X}$ $\sim$ 3.2$\times$10$^{34}$ erg
s$^{-1}$; Vink 2008) for the same energy range. This suggests that a
distance around 7 kpc for G337.8$-$0.1 is most probable. 

In summary, in this work we reported the first detection at X-ray
wavelengths of the SNR G337.8$-$0.1. The radio and X-ray properties
show that the object displays the three basics attributes found in
mixed-morphology SNRs, i.e. centrally filled X-ray morphology with
lines in its spectra, a shell-like appearance at radio frequencies, and 
the absence of a prominent, central, compact source in radio and X-ray
energies. If the SNR is indeed propagating in a cloudy medium, the
interaction of the shock-front with the adjacent molecular material
could be responsible for the gamma-ray emission detected by the EGRET
telescope.   

The detailed properties of the SNR are still unknown because of the low
photon statistics in the XMM-{\it Newton} data. Future X-ray
observations with a higher spatial resolution and good photon
statistics like Chandra or Suzaku will reveal the relation between the
X-ray intensity peak and the radio emission. 


\begin{acknowledgements} 
We thank the anonymous referee for her/his insightful comments and constructive 
suggestions that lead to an improved manuscript. The authors acknowledge 
support by DGI of the Spanish Ministerio de Educaci\'on y Ciencia under grants 
AYA2007-68034-C03-02, FEDER funds and Plan Andaluz de Investigaci\'on Desarrollo 
e Innovaci\'on (PAIDI) of Junta de Andaluc\'{\i}a as research group FQM322. 
J.F.A.C. is research of the Consejo Nacional de Investigaciones Cient\'ificas y
Tecnol\'ogicas (CONICET), Argentina. 

\end{acknowledgements}


\begin{thebibliography}{}

\bibitem[Anders \& Grevesse, 1989]{anders89} Anders, E., \& Grevesse, N.\ 1989, \gca, 53, 197.
\bibitem[Arnaud, 1996]{arnaud96} Arnaud, K.~A.\ 1996, ASP Conf.~Ser.~101: Astronomical Data Analysis Software and Systems V, 101, 17.
\bibitem[Brickhouse, 2003]{brick03} Brickhouse, N.~S.\ 2003, IAUJD, 17, 23.
\bibitem[Caswell et al., 1975]{caswell75} Caswell, J.~L., et al.\ 1975, \aap, 45, 239 



\bibitem[Cowie \& McKee, 1977]{cowie77} Cowie, L.~L., \& McKee, C.~F.\ 1977, \apj, 211, 135 
\bibitem[Cox et al., 1999]{cox99} Cox, D.~P., et al.\ 1999, \apj, 524, 179 

\bibitem[De Luca \& Molendi, 2004]{deluca04} De Luca, A., \& Molendi, S.\ 2004, \aap, 419, 837 
\bibitem[Dame, 1986]{dame86} Dame, T.M., Elmegreen, B.G., Cohen, R.S., Thaddeus, P., 1986, ApJ 305, 892
\bibitem[De Luca \& Molendi, 2004]{DeLuca04} De Luca, A. \& Molendi, S.\ 2004, 419, 837
\bibitem[Dickey \& Lockman, 1990]{Dickey90} Dickey, J.~M., Lockman, F.~J., ARA\&A 1990, 28, 215. 
\bibitem[Green et al., 1997]{green97} Green, A.J., Frail, D.A., Goss, W.M., \& Otrupcek, R. 1997, AJ, 114, 2058
\bibitem[Harrus et al., 1997]{harrus97} Harrus, I.~M., 1997, \apj, 488, 781 
\bibitem[Harrus et al., 2001]{harrus01} Harrus, I.~M., et al., 2001, \apj, 552, 614 
\bibitem[Hartman et al., 1999]{hatman99} Hartman, R.~C., et al.\ 1999, \apjs, 123, 79 
\bibitem[Hughes et al., 1998]{hughes98} Hughes, J.~P., Hayashi, I., \& Koyama, K.\ 1998, \apj, 505, 732 
\bibitem[Jones et al., 1993]{jones93} Jones, L.~R., Smith, A., \& Angelini, L.\ 1993, \mnras, 265, 631 
\bibitem[Koralesky et al., 1998]{koralevsky98} Koralesky, B., et al. \ 1998, \aj, 116, 1323 
\bibitem[Mazzotta et al., 1998]{mazzotta98} Mazzotta, P., Mazzitelli, G., Colafrancesco, S., \& Vittorio, N.\ 1998, \aaps, 133, 403 
\bibitem[Morrison \& McCammon, 1983]{morri83} Morrison, R., \& McCammon, D.\ 1983, \apj, 270, 119.
\bibitem[Press et al., 1992]{press92} Press, W.~H., Teukolsky, S.~A., Vetterling, W.~T., \& Flannery, B.~P.\ 1992, Cambridge: University Press, |c1992, 2nd ed.
\bibitem[Rho \& Petre, 1996]{rho96} Rho, J.-H., \& Petre, R.\ 1996, \apj, 467, 698 
\bibitem[Rho \& Petre, 1998]{rho98} Rho, J., \& Petre, R.\ 1998, \apjl, 503, L167 
\bibitem[Rho \& Borkowski, 2002]{rho02} Rho, J., \& Borkowski, K.~J.\ 2002, \apj, 575, 201 
\bibitem[Safi-Harb et al., 2005]{safi05} Safi-Harb, S., et al.\ 2005, \apj, 618, 321 
\bibitem[Shaver \& Goss, 1970]{shaver70} Shaver, P.~A., \& Goss, W.~M.\ 1970, Australian Journal of Physics Astrophysical Supplement, 17, 133 
\bibitem[Shelton et al., 1999]{shelton99} Shelton, R.~L., et al.\ 1999, \apj, 524, 192 


\bibitem[Slane et al., 2002]{slane02} Slane, P., Smith, R.~K., Hughes, J.~P., \& Petre, R.\ 2002, \apj, 564, 284 
\bibitem[Str{\"u}der et al., 2001]{stru01} Str{\"u}der, L., et al.\ 2001, \aap, 365, L18.
\bibitem[Turner, 2001]{tur01} Turner, M. J. L., et al. 2001, A\&A, 365, L27
\bibitem[Vink, 2004]{vink04} Vink, J.\ 2004, \apj, 604, 693 
\bibitem[White \& Long, 1991]{white91} White, R.~L., \& Long, K.~S.\ 1991, \apj, 373, 543 
\bibitem[Whiteoak \& Green, 1996]{whit96} Whiteoak, J.~B.~Z.~\& Green, A.~J. 1996, A\&A, 118, 329.



\end{thebibliography}
\end{document}